\documentclass[journal=jpccck,manuscript=article]{achemso}

\usepackage[version=3]{mhchem} % Formula subscripts using \ce{}
\usepackage[T1]{fontenc}       % Use modern font encodings
\usepackage{graphicx}          % grafiken einbinden
\usepackage{amsmath}
\setkeys{acs}{usetitle = true}

\usepackage[english]{babel}
\usepackage[latin9]{inputenc}
\usepackage{amsthm}
\usepackage{amssymb}
\usepackage{esint}
\usepackage{upgreek}
\usepackage{siunitx}

\makeatletter
\@ifundefined{textcolor}{}
{%
 \definecolor{BLACK}{gray}{0}
 \definecolor{WHITE}{gray}{1}
 \definecolor{RED}{rgb}{1,0,0}
 \definecolor{GREEN}{rgb}{0,1,0}
 \definecolor{BLUE}{rgb}{0,0,1}
 \definecolor{CYAN}{cmyk}{1,0,0,0}
 \definecolor{MAGENTA}{cmyk}{0,1,0,0}
 \definecolor{YELLOW}{cmyk}{0,0,1,0}
}

\usepackage{soul}
\usepackage{xspace}
\usepackage{braket}
\usepackage[backref=true,
bookmarksnumbered=true,
bookmarks=true,
bookmarksopen=true,
colorlinks=true,
citecolor=blue,
linkcolor=blue,
anchorcolor=green,
urlcolor=blue,unicode=false]{hyperref}

\let\baraccent=\= % rename builtin command \= to \baraccent
\renewcommand{\=}[1]{\stackrel{#1}{=}} % for putting numbers above =

\newcommand{\molf}{Ethyl-Diaminodicyanoquinone\xspace} % molecule name
\newcommand{\mol}{Ethyl-DADQ\xspace} % molecule name
\newcommand{\posg}{imidazolidine group\xspace} % molecule name
\newcommand{\negg}{dicyanomethylene group\xspace} % molecule name
\newcommand{\ion}{ionization level\xspace}
\newcommand{\aff}{affinity level\xspace}
\makeatother

\title{Resolution of intramolecular dipoles and push-back effect of individual
  molecules\\ on a metal surface}

\author{Sergey Trishin}
\affiliation{\mbox{Fachbereich Physik, Freie Universit\"at Berlin, 14195 Berlin, Germany}}
\altaffiliation{These authors contributed equally.}

\author{Tobias M\"uller}
\affiliation{Interdisciplinary Center for Molecular Materials (ICMM) and
  Computer Chemistry Center (CCC), Friedrich-Alexander-Universit\"at\\
  Erlangen-N\"urnberg, 91052 Erlangen, Germany}
\altaffiliation{These authors contributed equally.}

\author{Daniela Rolf}
\affiliation{\mbox{Fachbereich Physik, Freie Universit\"at Berlin, 14195 Berlin, Germany}}

\author{Christian Lotze}
\affiliation{\mbox{Fachbereich Physik, Freie Universit\"at Berlin, 14195 Berlin, Germany}}

\author{Philipp Rietsch}
\affiliation{\mbox{Institut f\"ur Chemie und Biochemie, Freie Universit\"at Berlin, 14195 Berlin, Germany}}

\author{Siegfried Eigler}
\affiliation{\mbox{Institut f\"ur Chemie und Biochemie, Freie Universit\"at Berlin, 14195 Berlin, Germany}}

\author{Bernd Meyer}
\affiliation{Interdisciplinary Center for Molecular Materials (ICMM) and
  Computer Chemistry Center (CCC), Friedrich-Alexander-Universit\"at\\
  Erlangen-N\"urnberg, 91052 Erlangen, Germany}
\email{bernd.meyer@chemie.uni-erlangen.de}

\author{Katharina J. Franke}
\affiliation{\mbox{Fachbereich Physik, Freie Universit\"at Berlin, 14195 Berlin, Germany}}
\email{franke@physik.fu-berlin.de}
\begin{document}

\maketitle

\begin{abstract}
Molecules consisting of a donor and an acceptor moiety can exhibit large
intrinsic dipole moments. Upon deposition on a metal surface, the dipole
may be effectively screened and the charge distribution altered due to
hybridization with substrate electronic states. Here, we deposit \molf
molecules, which exhibit a large dipole moment in gas phase, on a Au(111)
surface. Employing a combination of scanning tunneling microscopy and
non-contact atomic force microscopy, we find that a significant dipole
moment persists in the flat-lying molecules. Density-functional theory
calculations reveal that the dipole moment is even increased on the metal
substrate as compared to the gas phase. We also show that the local contact
potential across the molecular islands is decreased by several tens of meV
with respect to the bare metal. We explain this by the induced charge-density
redistribution due to the adsorbed molecules, which confine the substrate's
wavefunction at the interface. Our local measurements provide direct evidence
of this so-called push-back or cushion effect at the scale of individual
molecules.
\end{abstract}

%\pacs{75.75.-c,74.20.-z,75.70.Tj,73.63.Nm}

%------------------------------------------------------------------------------
\section{Introduction}
%------------------------------------------------------------------------------

Donor-acceptor molecules exhibit two charge-separated moieties, which make
them interesting for applications in molecular electronics
\cite{aviram1,metzger1} and optoelectronic devices \cite{bergkamp1}. When
adsorbed on a surface, the intramolecular charge distribution may change
drastically \cite{meier1, Queck2019}. In particular, the adsorption on metal
surfaces usually leads to hybridization, charge transfer and screening, which
alter the molecular properties \cite{Ishii1999, Braun2009, Torrente2008}.
To reduce hybridization and screening from the substrate, thin insulating
layers are often introduced at the interface to a metal substrate
\cite{repp1, Qiu2004, Kahle2012, Schulz2013, Krane2018, mohn1}. However,
some applications may require adsorption directly onto metal surfaces.
Therefore, finding a suitable donor--acceptor molecule, which preserves
its properties upon adsorption on a metal surface, is of great interest.

Another interesting aspect of molecular adsorbate layers on metals is their
potential to tune charge injection barriers at organo-metallic interfaces
\cite{kahn1,Koch2005}.
Typically, a molecular adsorbate layer suppresses the decay of the
metallic states into the vacuum. This results in a reduction of the
spill-out of the electron charge. The induced electric dipole perpendicular
to the surface leads to a lower workfunction. The reduced workfunction can
be easily detected in photoelectron spectroscopy, with the phenomenon being
termed push-back or cushion effect \cite{Ishii1999}.

The resolution of intramolecular dipoles and the push-back effect on the
single-molecule scale requires a technique, which probes the structure and
charge distributions with atomic resolution. Kelvin probe force microscopy
(KPFM) measures the local contact potential difference (LCPD) between a
conducting tip of an atomic force microscope (AFM) and a sample
\cite{Nonnenmacher1991}.
Since its invention, KPFM has quickly evolved as a highly versatile tool
to map charge distributions and workfunction changes with nanoscale
resolution, most commonly applied to metallic and semiconducting structures
\cite{melitz1}. With the development of qPlus sensors, KPFM could be
combined with scanning tunneling microscopy (STM). This combination led to
the seminal mapping of charge distributions at the atomic scale
\cite{Gross2009} and to a number of important insights into intramolecular
charge distributions \cite{mohn1, schuler1, albrecht1, meier1, Albrecht2015,
krull2018, Queck2019}.
However, a quantitative determination of the charge distribution within
molecules and molecular layers remains challenging and it is important to
consider that the LCPD value is detected at a certain tip height
\cite{schuler1}.

\begin{figure}[!t]
\includegraphics[width=0.96\textwidth]{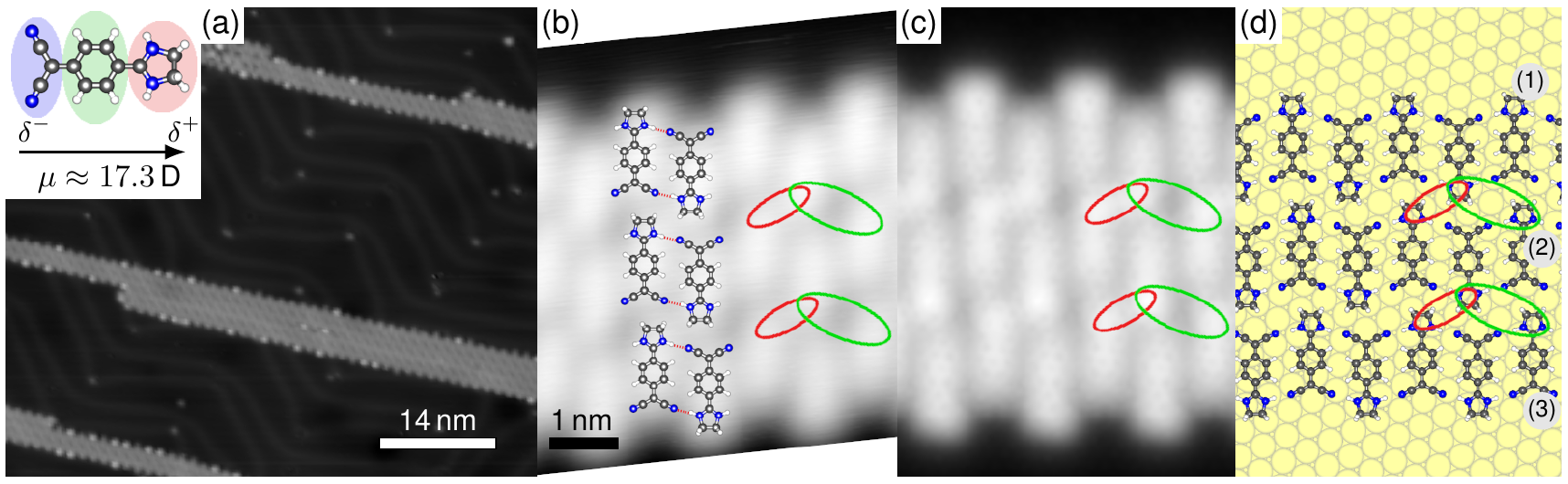}
\caption{\label{F:stm}
  (a) STM topography of \mol adsorbed on a Au(111) surface. The topography was
  recorded at \SI{1}{V}, \SI{100}{\pA}. Inset: Structure of the \mol molecule.
  It consists of a benzene ring (green) with a dicyanomethylene (tail, purple)
  and a imidazolidine (head, red) group bound to it. The different electron
  affinities of the head and tail group lead to a large dipole moment of about
  17.3\;Debye in the gas phase.
  (b) STM topography of a chain of \mol. The superimposed structure model
  (optimized by DFT) indicates the proposed orientation of the molecules.
  The red dotted lines represent CN$\cdots$HN hydrogen bonds. The topography
  was recorded at \SI{2}{V}, \SI{306}{\pA}. (c) Simulated STM topography
  (\SI{1}{V}, isodensity 2$\cdot$10$^{-4}$\,$e$/{\AA}$^3$) of a three-molecule
  wide chain on a Au(111) substrate, calculated for the relaxed structure
  shown in (d). Red and green ellipses indicate the asymmetric head-to-tail
  contact between rows of the chain, which is visible in STM as alternating 
  brighter and darker dots.}
\end{figure}

Here, we chose Ethyl-Diaminodicyanoquinone (\mol) as a potential candidate
to map intramolecular charges as well as the push-back effect on a Au(111)
substrate. These molecules have a large dipole moment due to their
donor--acceptor units, and they have shown recently to be highly fluorescent
\cite{rietsch1}. They consist of a quinone moiety with a dicyanomethylene
and an imidazolidine group bound to it on opposite sites. Figure\,\ref{F:stm}a
(inset) shows the optimized gas-phase structure of the molecule. The two
end groups have different electron affinities, which results in a large
lateral dipole moment of about 17.3\,Debye. Apart from two hydrogen atoms
of the \posg that stick out of the molecular plane, the molecule is flat
and thus ideally suited for obtaining an in-plane dipole on a surface.

%------------------------------------------------------------------------------
\section{Methods}
%------------------------------------------------------------------------------

We have performed combined STM/AFM experiments probing \mol on a Au(111)
single-crystal surface. The Au surface was cleaned by repeated sputter/anneal
cycles until a clean, atomically flat termination was obtained. \mol molecules
were evaporated at 500\,K onto the clean sample held at room temperature.
The experiments were carried out at a base temperature of 4.8\,K.
Differential-conductance spectra were taken with a lock-in amplifier at a
modulation frequency of 909\,Hz. A qPlus sensor \cite{giessibl1} was used
for combined STM/AFM measurements.

The LCPD is obtained by recording the
frequency shift while the bias voltage is ramped \cite{Gross2009}. Due to the 
quadratic dependence of the frequency shift on the bias voltage, the frequency
shift $\Delta f$ as a function of the bias voltage $V$ is an inverse parabola:
$$\Delta f = \dfrac{-f_0}{2k}\dfrac{\partial F}{\partial z}\;,\quad
F_{\rm el}=\dfrac{V^2}{2} \dfrac{\partial C}{\partial z}.$$
$f_0$ and $k$ are the resonance frequency and the spring constant of the
oscillating tip, $F$ and $F_{\rm el}$ are the total and the electrostatic
contribution of the force acting on the tip, and $C$ is the capacitance
of the junction. By a parabolic fit of the individual $\Delta f(V)$ curves,
one can extract the voltage at the maximum of the curve, which is needed to
compensate for the electrostatic forces due to charges at the interface
\cite{Gross2009}.

Periodic density-functional theory (DFT) calculations were performed with
the periodic plane-wave code \texttt{PWscf} of the Quantum Espresso software
package \cite{Giannozzi2009}, using the PBE exchange-correlation functional
of Perdew, Burke and Ernzerhoff \cite{Perdew1996}, Vanderbilt ultrasoft
pseudopotentials \cite{Vanderbilt1990}, and a plane-wave basis set with an
energy cutoff of 30\,Ry. Dispersion corrections to PBE energies and forces
were included by our recently introduced D3$^{\rm surf}$ scheme.
D3$^{\rm surf}$ is an extension of the original D3 method proposed by Grimme
et al. \cite{Grimme2010}, in which the parameter set of coordination-dependent
$C_6$ coefficients is extended by additional values for atoms at higher
coordination numbers \cite{Kachel2020, Wolfram2022}.

The Au(111) surface was represented by periodically repeated slabs with a
thickness of four atomic layers. The calculated PBE+D3$^{\rm surf}$ Au bulk
lattice constant of 4.118\,{\AA} was used for the lateral extension of the
slabs. The structure search was carried out with various surface-unit-cell
sizes representing different surface coverages and adsorbate structures.
Unit cells of the slabs contained up to 456 gold atoms. In the geometry
optimizations, the bottom two atomic layers were kept fixed, while the
upper two layers and the adsorbed molecules were relaxed. The density of
the Monkhorst-Pack k-point mesh for Brillouin zone sampling was larger or
equivalent to a (24,24,1) mesh for the primitive surface unit cell. STM
images were simulated using the Tersoff-Hamann approximation \cite{TH85}.

%------------------------------------------------------------------------------
\section{Results and Discussion}
%------------------------------------------------------------------------------
\subsection{Structure of Ethyl-DADQ on Au(111)}
%------------------------------------------------------------------------------

Ethyl-DADQ molecules deposited onto a Au(111) surface at room temperature
self-assemble into straight chains across the whole surface (STM image in
Fig.\,\ref{F:stm}a). The chains do not affect the Au(111) reconstruction lines
and do not show preferred adsorption directions with respect to the symmetry
of the Au surface, indicating a rather weak chemical bonding to the surface
\cite{vanderlit1}. The widths of the chains can be tuned by varying the
molecular coverage. A close-up view on one of these chains is shown in
Fig.\,\ref{F:stm}b. Each individual molecule is imaged with an almost oval
shape with one termination being slightly wider than the other one.
Across the rows of the chains, the molecules are arranged head to tail. The
neighboring molecules within a row are oriented in the opposite direction
and shifted with respect to each other. The suggested model is indicated by
a superimposed molecular structure onto the STM image. Such an alignment is
favored by the formation of CN$\cdots$HN hydrogen bonds (indicated by red
dotted lines) and electrostatic interactions due to the anti-parallel
alignment of the dipoles.

To rationalize this model, we performed a systematic structure search using
density-functional theory (DFT) calculations. To arrive at a realistic
structure, we first determined the preferred adsorption site and molecular
orientation of a single Ethyl-DADQ molecule in a large surface unit cell.
We find that the preference of the adsorption site and molecular orientation
is not very pronounced, as the energy cost to shift and tilt the molecule out
of the preferred site and orientation is rather small. Such a shallow
adsorption energy landscape indicates that the adsorption is mainly driven by
isotropic van-der-Waals forces and electrostatic interactions. Nonetheless,
the adsorption energy of 1.96\,eV is quite large (see Table\,\ref{T:Eads}).

Next, we constructed molecular dimers and trimers on the surface to determine
the preferred alignment and the equilibrium distances of the molecules.
Interestingly, we find a huge energy gain of 0.51\,eV per molecule for
an antiparallel orientation of the Ethyl-DADQs on the surface (see
Table\,\ref{T:Eads}). This structure is mainly stabilized by CN$\cdots$HN
hydrogen bonds, which leads to the characteristic parallel displacement of
the cyano and the imidazolidine groups between neighboring molecules. As
the antiparallel alignment entails the largest contribution to the bonding
energy, extended molecular structures are expected to be quasi
one-dimensional, in agreement with the observation of long chains in
experiment.

\begin{table}[!b]
\centering
\def\tabcolsep{10pt}
\def\arraystretch{1.2}
\begin{tabular}{ccc}
\hline\hline
Structure & $E_{\rm ads}$ (eV) & $\Delta E_{\rm ads}$ (eV) \\
\hline
Monomer      & -1.96 &  $-$  \\
Dimer        & -2.47 & -0.51 \\
Trimer       & -2.57 & -0.10 \\
Single chain & -2.77 & -0.20 \\
Double chain & -2.81 & -0.04 \\
Triple chain & -2.83 & -0.02 \\
Monolayer    & -2.87 & -0.04 \\
\hline\hline
\end{tabular}
\caption{\label{T:Eads}
  Adsorption energies $E_{\rm ads}$ per Ethyl-DADQ molecule for the
  energetically most favorable configuration of different molecular
  assemblies on Au(111) with increasing surface coverage. The change
  of the adsorption energy from coverage to coverage is given by
  $\Delta E_{\rm ads}$\,.}
\end{table}

To capture the structure of extended chains on the Au(111) surface, we
constructed commensurate models of chains with increasing width. In doing so,
we made sure that the molecular distances of the dimers and trimers as well
as their orientation and relative position with respect to the substrate is
maintained as closely as possible. This was fulfilled by orienting rows of
molecules along the crystallographic $[23\bar{5}]$--direction (with
periodicity of $\tfrac{1}{2}[23\bar{5}]$) and by stacking them asymmetrically
together.
Consecutive rows are slightly shifted so that in the head-to-tail alignment
of molecules between neighboring rows, the imidazolidine group is closer to
one of the two CN groups than the other.
This is shown in Fig.\,\ref{F:stm}d for the case of a three-molecule
wide chain. The symmetry breaking is rooted in optimizing the molecular
head-to-tail contact and the relativ position of the molecules with respect
to the substrate.
Finally, two-dimensional commensurate models for describing inner parts of
wider chains were built by connecting the sides of a two-molecule wide chain
by a corresponding unit-cell vector. This was done again in such a way that
all structural elements and the asymmetric head-to-tail contacts between rows
are reproduced as found for the chains with finite width.

The structure obtained by this systematic search is in very good agreement
with the experimental results. Direct evidence of the match can be obtained
from a simulated STM image shown in Fig.\,\ref{F:stm}c. This captures very
well the experimentally observed asymmetry of the individual molecules.
In addition, fine details such as the alternating distance between
imidazolidine head and CN groups of molecules in neighboring rows are resolved
that can equally be found in experiment and are reflected in a periodic line
of brighter and darker dots between the molecules, see Fig.\,\ref{F:stm}b-d.

\begin{figure}[!b]
\includegraphics[width=0.48\textwidth]{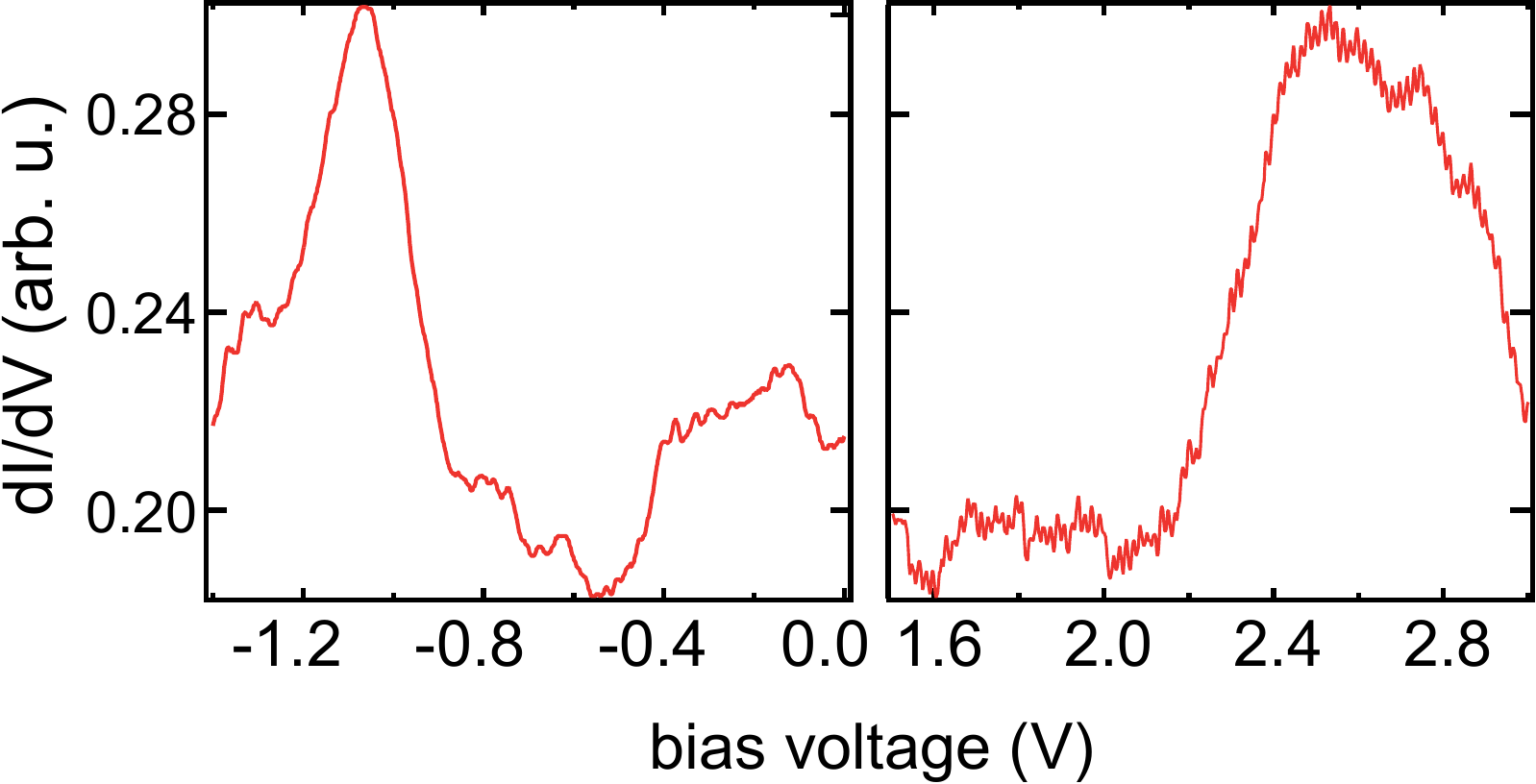}
\caption{\label{F:spectra}
  d$I$/d$V$ spectra recorded on a molecule. Resonances can be seen around
  \SI{-1}{V} (left side, attributed to the \ion) and \SI{2.5}{V}
  (right side, attributed to the \aff).  The left spectrum was recorded at
  a setpoint of \SI{1}{V}, \SI{100}{\pA}, the right spectrum was recorded
  in the constant current mode at \SI{100}{\pA}. The modulation amplitude
  was \SI{5}{mV} (left) and \SI{10}{mV} (right).}
\end{figure}

%------------------------------------------------------------------------------
\subsection{Charge distribution within Ethyl-DADQ on Au(111)}
%------------------------------------------------------------------------------

To learn about the electronic properties of the molecules in this structure,
we performed d$I$/d$V$ spectroscopy. Figure\,\ref{F:spectra} shows a constant
height (left) and a constant-current (right) d$I$/d$V$ spectrum recorded on a
molecule. Two  resonances can be observed, one located at around $-1$\,V and
the other at around 2.5\,V. We attribute these resonances to tunneling into
the ionization and from the affinity level of the molecules, respectively.
The large noise signal in the spectra at positive bias voltages indicates an
instability of the molecules at these bias voltages. The width of the positive
ion resonance amounts to $\sim$160\,meV, which is narrower than resonances
from other extended organic molecules on Au(111). The narrow linewidth points
to a weaker coupling to the surface than typically observed \cite{Torrente2008}.
The negative ion resonance appears broader, which probably results from
excitation of vibronic states \cite{Qiu2004} and vibration-assisted tunneling
processes \cite{pavlicek1, Reecht2020}.

\begin{figure}[!t]
\includegraphics[width=0.96\textwidth]{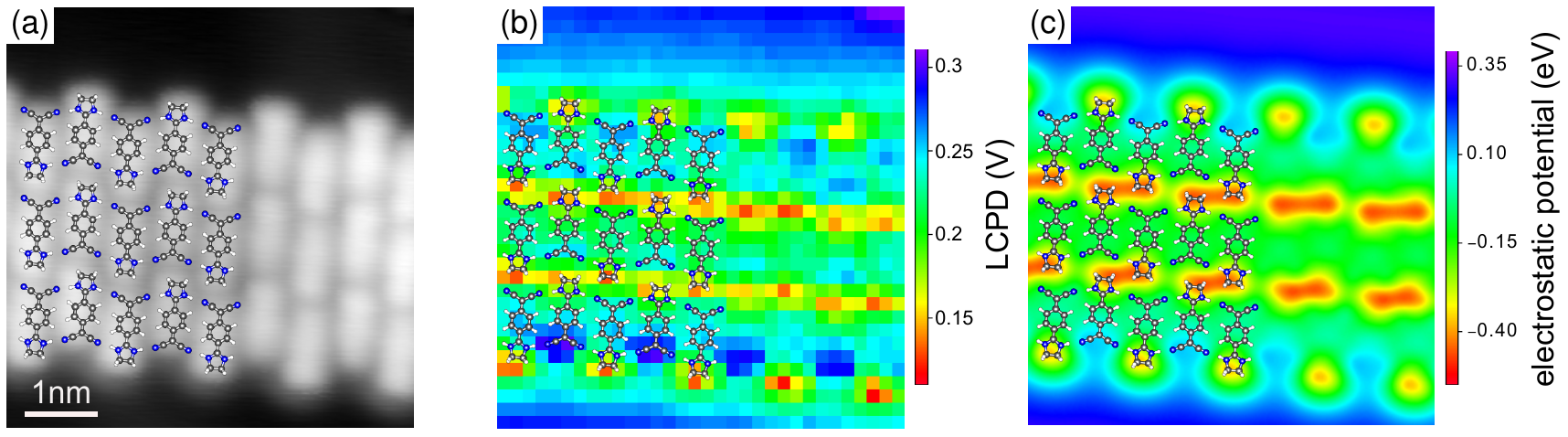}
\caption{\label{F:lcpdmap}
  (a) STM topography of a three-molecule wide chain, recorded at
  \SI{1.5}{V}, \SI{130}{\pA}. (b) LCPD values extracted from a grid of
  spectra along the same area as in (a). The spectra were recorded at
  a setpoint of \SI{1.5}{V}, \SI{130}{\pA}. Additionally, the tip was
  approached by 0.8\,{\AA} towards the molecule for each spectrum.
  (c) Calculated electrostatic potential of a three-molecule wide chain
  at a height of 7.1\,{\AA} above the surface. The planar averaged
  potential is set to zero.}
\end{figure}

To find out if the dipole is preserved upon adsorption, we measured the
LCPD signal along a three-molecule wide chain as shown by STM in
Fig.\,\ref{F:lcpdmap}a. We find an overall decrease of the LCPD values on
the molecules as compared to the bare substrate (Fig.\,\ref{F:lcpdmap}b).
Additionally, the signal is modulated within the molecular island. The
superimposed structural model obtained from the simultaneously recorded
STM image shows that the regions of lowest LCPD correspond to the \posg 
of the \mol.

\begin{figure}[!t]
\includegraphics[width=0.48\textwidth]{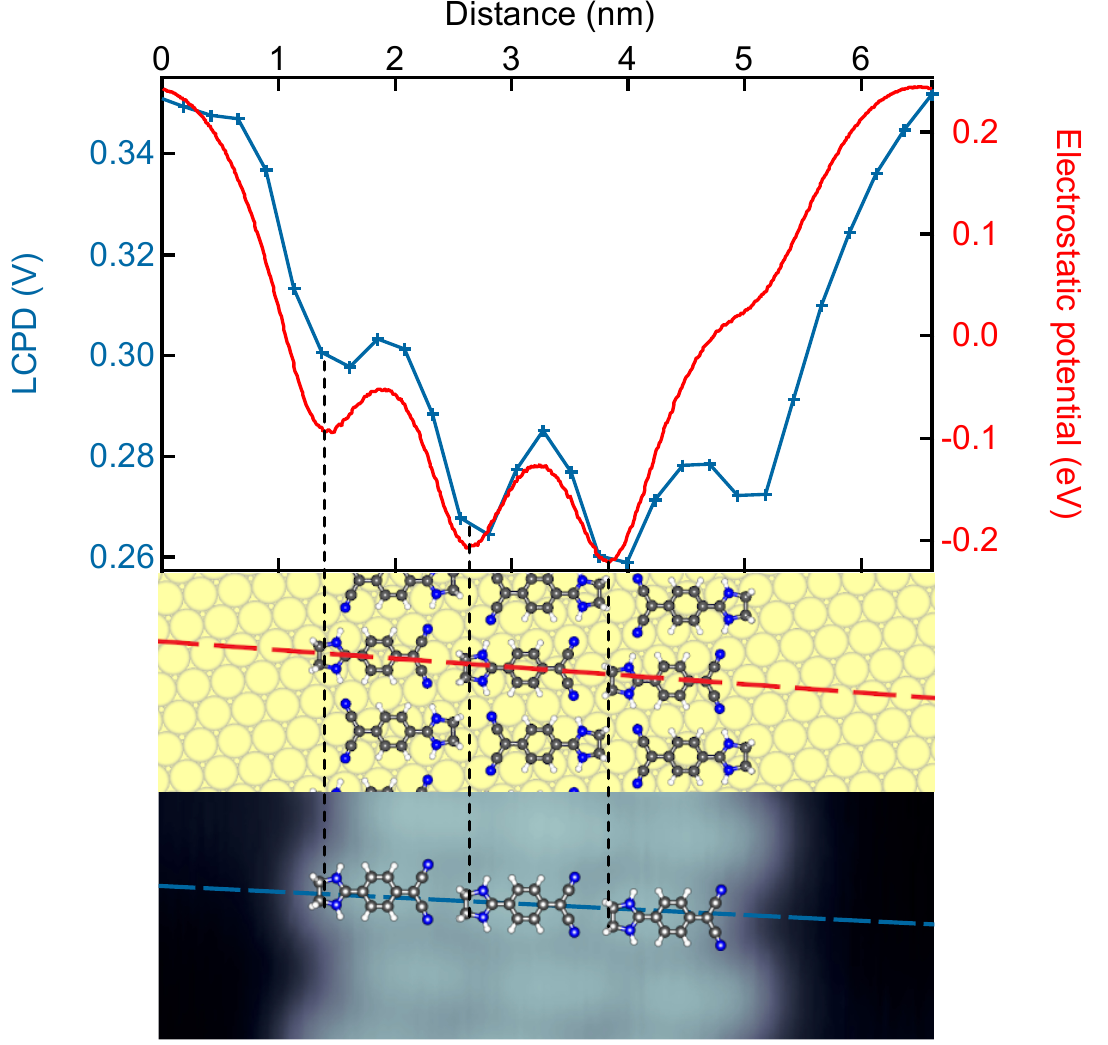}
\caption{\label{F:lcpd}
  Plot of the LCPD value (blue solid line) measured across a molecular chain
  as indicated by the blue dashed line in the STM topography below the graph.
  The spectra were recorded at a setpoint of \SI{2}{V}, \SI{300}{\pA} and the
  STM topography was recorded at a setpoint of \SI{1}{V}, \SI{172}{\pA}.
  The calculated electrostatic potential across the molecular chain at a
  height of 9.3\,{\AA} above the surface is shown in red. The planar averaged
  potential is set to zero. The vertical black dashed lines indicate the
  position of the \posg of each individual molecule in the chain.}
\end{figure}

A better perception of the modulation along the molecules can be gained
from a line profile of the LCPD signal across a three-molecule wide island
(see Fig.\,\ref{F:lcpd}, blue curve). In agreement with the map, we observe
the overall reduction of the LCPD above the molecular layer as well as the
additional oscillation on the molecular scale. As illustrated by the
comparison of the LCPD line and the topographic image, the oscillation can
be directly linked to the locations along the individual molecules. The
lowest LCPD value is found at the \posg of the \mol, whereas the largest
value is found toward the \negg, but not exactly on top of it as may be
expected from the dipolar charge distribution of the gas-phase molecule with
positive charge at the \posg and negative charge at the \negg. This is due
the fact that the measured LCPD value corresponds to the electrostatic
potential at the tip height above the island.
Furthermore, we note that the values at the edges of the chain are
affected by the overall decrease of the LCPD and, additionally on the right
side, by the neighboring molecules sticking out of the chain structure.
However, from the LCPD modulations, we can conclude that the \mol molecule
preserves a finite dipole moment even on the metallic Au(111) surface.

To corroborate this interpretation, we calculated the charge distribution
of the surface-adsorbed \mol molecules. We started with a simple Bader
charge analysis which we benchmark against the gas-phase molecule.
As a test to see how well Bader charges reproduce the dipolar properties of
a molecule, we used the atomic Bader charges of the gas-phase \mol molecule
and calculated the molecular dipole moment. We obtain a value of 25.5\,Debye,
which is in reasonably good agreement with the true dipole moment of
17.3\,Debye as determined from the full charge distribution within the
molecule. We can therefore expect that Bader charges capture well changes
in intramolecular charge distributions and molecular dipole moments if
molecules are transferred into a different environment.

\begin{table}[!b]
\centering
\def\tabcolsep{6pt}
\def\arraystretch{1.2}
\begin{tabular}{ccccc}
\hline\hline
Structure & cyano & benzene & imid & $\Delta q$ \\[-4pt]
          & unit  & unit    & unit &  \\
\hline
Gas phase          & -0.51 & +0.31 & +0.20 &  $-$  \\
Monomer            & -0.54 & +0.29 & +0.40 & +0.15 \\
Single chain       & -0.64 & +0.24 & +0.48 & +0.08 \\
Triple chain   (1) & -0.64 & +0.23 & +0.47 & +0.06 \\
\hspace*{64pt} (2) & -0.67 & +0.24 & +0.47 & +0.04 \\
\hspace*{64pt} (3) & -0.67 & +0.24 & +0.48 & +0.05 \\
Monolayer          & -0.67 & +0.24 & +0.47 & +0.04 \\
\hline\hline
\end{tabular}
\caption{\label{T:bader}
  Bader charges of the three subunits of Ethyl-DADQ molecules (see inset of
  Fig.\,\ref{F:stm}a) in the gas phase and adsorbed on the Au(111) surface
  for different coverages and structures. For the 3-molecule wide chain, the
  Bader charges of the three nonequivalent molecules (1) to (3) as indicated
  in Fig.\,\ref{F:stm}d are given separately. $\Delta q$ is the charge
  transfer from the molecules to the Au substrate.}
\end{table}

For analyzing the variation in charge distribution upon adsorption on the
surface and within different arrangements, we divided the \mol molecule
into three units: the dicyanomethylene tail, the central benzene ring, and
the imidazolidine head. The Bader charges of the atoms within each unit
are added and the total Bader charges of the three units are reported in
Table\,\ref{T:bader}.
As expected, the results show an accumulation of electrons on the cyano groups
and an electron depletion on the benzene ring and the imidazolidine unit.
Interestingly, upon adsorption on the Au(111) surface, the dipolar character
is not reduced but even enhanced, with a small net transfer of electrons to
the Au substrate.

\begin{figure}[!t]
\includegraphics[width=0.48\textwidth]{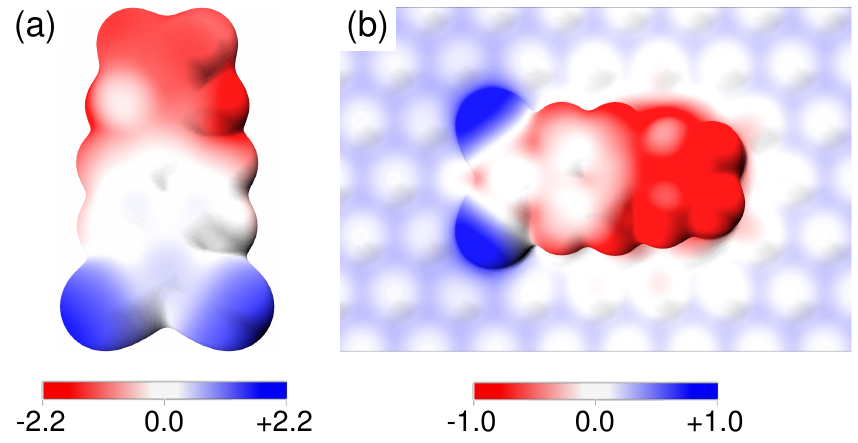}
\caption{\label{F:mep}
  Molecular electrostatic potential drawn on an isosurface of the electron
  density with isovalue of 0.002\,a.u.\ for (a) a Ethyl-DADQ molecule in
  the gas phase and (b) for an adsorbed monomer on Au(111). The color-coded
  potential values are given in eV with respect to the vacuum level. Red
  areas are attractive for electrons, blue areas are repulsive.}
\end{figure}

For better comparison with experiment, we calculated the electrostatic
potential above the surface which can be directly compared to the LCPD
measurements.
We first note that the distribution of the electrostatic potential supports
the conclusion from the Bader charge analysis. Figure\,\ref{F:mep}a shows
the molecular electrostatic potential on the van der Waals surface of a
gas-phase Ethyl-DADQ molecule. As expected, the potential is repulsive for
electrons above the negatively charged cyano groups, whereas the positive
charge on the imidazolidine unit creates an attractive potential. After
adsorption on the Au(111) surface, this characteristic distribution of
the molecular electrostatic potential is maintained, see Fig.\,\ref{F:mep}b,
which indicates that the intramolecular charge transfer and thus the molecular
dipole moment are preserved.

Within a densely packed molecular layer, the charge distribution within
the individual molecules may be influenced by the local environment and
deviate from an isolated molecule. Furthermore, it is important to note that
the measurements probe the LCPD at a certain tip height above the molecular
layer. To account for these effects and for direct comparison with
experiment, we plot the two-dimensional map of the electrostatic potential
(Fig.\,\ref{F:lcpdmap}c) and a corresponding line profile
(Fig.\,\ref{F:lcpd}, red line) across a three-molecule wide chain. While
the main character of the variation of the electrostatic potential is
preserved along the individual molecules, it is notable that the largest
LCPD is not located precisely at the \negg, but affected by the \posg of
the neighboring molecule and thus slightly shifted toward the molecule's
center. These details are in very good agreement with the variation of the
LCPD signal. The calculations thus corroborate the dipolar character of the
\mol molecules on the Au(111) surface and their persistence in the
densely-packed islands.

\begin{figure}[!t]
\includegraphics[width=0.48\textwidth]{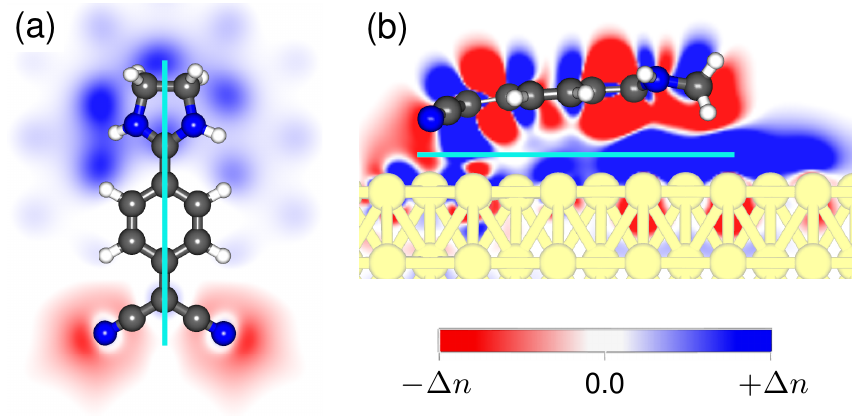}
\caption{\label{F:dendiff}
  Electron density difference plot for a Ethyl-DADQ monomer on Au(111) in
  (a) a plane 1\,{\AA} above the surface atoms
  ($\Delta n$\,=\,1.5$\cdot$10$^{-2}$\,$e$/{\AA}$^3$)
  and (b) a plane perpendicular to the surface through the central axis
  of the molecule
  ($\Delta n$\,=\,1.5$\cdot$10$^{-3}$\,$e$/{\AA}$^3$).
  The positions of the planes are indicated by cyan lines. The orientation
  of the molecules is the same as in Fig.\,\ref{F:mep}.
  Blue areas show electron accumulation, red areas electron depletion.}
\end{figure}

In agreement with experiment, the calculations also show the overall
reduction of the electrostatic potential above the molecules compared
to the plain Au(111) surface.
To unravel the origin of this, we analyzed the induced charge density
redistributions. They are calculated by subtracting the electron density
of the molecule and the surface determined by separate calculations for
the two fragments from the electron density of the combined structure.
Fig.\,\ref{F:dendiff}a shows the electron density difference in a plane
1\,{\AA} above the surface. We find a significant electron accumulation
underneath the imidazolidine unit whereas electron depletion is observed
underneath the cyano groups. This is direct evidence that the dipole of
the adsorbed molecule induces a compensating mirror dipole in the metal
substrate.
A cut through the electron density difference perpendicular to the surface
through the central axis of the molecule (see Fig.\,\ref{F:dendiff}b)
visualizes the induced electron density redistribution due to the presence
of the molecules. Close to the imidazolidine group a zone of electron
depletion is visible, which is in agreement with the more positive Bader
charge upon adsorption, see above. However, in a large area between the
molecule and the surface, the electron density is increased, which
demonstrates the compression (push-back) of the electron density of the
metal surface leaking into the vacuum, which is also termed ``cushion effect''
or ``pillow effect'' \cite{Ishii1999,kahn1,heimel1,koch1,schlesinger1}.
A consequence of this is the reduction on the local work function.

%------------------------------------------------------------------------------
\section{Conclusions}
%------------------------------------------------------------------------------

We investigated \molf molecules adsorbed on a Au(111) surface. By measuring
the local contact potential using a nc-AFM we observed that the intramolecular
dipole is preserved upon adsorption on the surface. Interestingly, our
calculations indicate that the dipole is even increased upon adsorption on the
metal surface. We also showed that the local work function is generally lower
on the molecular sites as compared to the bare metal substrate. This effect
originates from a local-scale push-back effect, where the molecules repel the
wave function of the substrate electrons, leaking out into the vacuum. Both
observations could be equally interesting for application in single-molecule
electronics and the energy alignment at metal--organic interfaces. 

%------------------------------------------------------------------------------
\section{Acknowledgements}
We acknowledge financial support by the Deutsche Forschungsgemeinschaft (DFG)
through SFB~951 ``Hybrid Inorganic/Organic Systems for Opto-Electronics''
(project number 182087777, project~A14) and SFB~953 ``Synthetic Carbon
Allotropes'' (project number 182849149, project~C1).

%------------------------------------------------------------------------------
%\bibliographystyle{achemso}
%\bibliography{bib}

\begin{mcitethebibliography}{38}
\providecommand*\natexlab[1]{#1}
\providecommand*\mciteSetBstSublistMode[1]{}
\providecommand*\mciteSetBstMaxWidthForm[2]{}
\providecommand*\mciteBstWouldAddEndPuncttrue
  {\def\EndOfBibitem{\unskip.}}
\providecommand*\mciteBstWouldAddEndPunctfalse
  {\let\EndOfBibitem\relax}
\providecommand*\mciteSetBstMidEndSepPunct[3]{}
\providecommand*\mciteSetBstSublistLabelBeginEnd[3]{}
\providecommand*\EndOfBibitem{}
\mciteSetBstSublistMode{f}
\mciteSetBstMaxWidthForm{subitem}{(\alph{mcitesubitemcount})}
\mciteSetBstSublistLabelBeginEnd
  {\mcitemaxwidthsubitemform\space}
  {\relax}
  {\relax}

\bibitem[Aviram and Ratner(1974)Aviram, and Ratner]{aviram1}
Aviram,~A.; Ratner,~M.~A. Molecular {Rectifiers}. \emph{Chem. Phys. Lett.}
  \textbf{1974}, \emph{29}, 277\relax
\mciteBstWouldAddEndPuncttrue
\mciteSetBstMidEndSepPunct{\mcitedefaultmidpunct}
{\mcitedefaultendpunct}{\mcitedefaultseppunct}\relax
\EndOfBibitem
\bibitem[Metzger(2003)]{metzger1}
Metzger,~R.~M. Unimolecular {Electrical} {Rectifiers}. \emph{Chem. Rev.}
  \textbf{2003}, \emph{103}, 3803\relax
\mciteBstWouldAddEndPuncttrue
\mciteSetBstMidEndSepPunct{\mcitedefaultmidpunct}
{\mcitedefaultendpunct}{\mcitedefaultseppunct}\relax
\EndOfBibitem
\bibitem[Bergkamp \latin{et~al.}(2015)Bergkamp, Decurtins, and Liu]{bergkamp1}
Bergkamp,~J.; Decurtins,~S.; Liu,~S.-X. Current {Advances} in {Fused}
  {Tetrathiafulvalene} {Donor}-{Acceptor} {Systems}. \emph{Chem. Soc. Rev.}
  \textbf{2015}, \emph{44}, 863\relax
\mciteBstWouldAddEndPuncttrue
\mciteSetBstMidEndSepPunct{\mcitedefaultmidpunct}
{\mcitedefaultendpunct}{\mcitedefaultseppunct}\relax
\EndOfBibitem
\bibitem[Meier \latin{et~al.}(2017)Meier, Pawlak, Kawai, Geng, Liu, Decurtins,
  Hapala, Baratoff, Liu, Jel\'{i}nek, Meyer, and Glatzel]{meier1}
Meier,~T.; Pawlak,~R.; Kawai,~S.; Geng,~Y.; Liu,~X.; Decurtins,~S.; Hapala,~P.;
  Baratoff,~A.; Liu,~S.-X.; Jel\'{i}nek,~P. \latin{et~al.}  Donor Acceptor
  Properties of a Single Molecule Altered by On Surface Complex Formation.
  \emph{ACS Nano} \textbf{2017}, \emph{11}, 8413--8420\relax
\mciteBstWouldAddEndPuncttrue
\mciteSetBstMidEndSepPunct{\mcitedefaultmidpunct}
{\mcitedefaultendpunct}{\mcitedefaultseppunct}\relax
\EndOfBibitem
\bibitem[Queck \latin{et~al.}(2019)Queck, Albrecht, Mutombo, Krejci,
  Jel\'{i}nek, McLean, and Repp]{Queck2019}
Queck,~F.; Albrecht,~F.; Mutombo,~P.; Krejci,~O.; Jel\'{i}nek,~P.; McLean,~A.;
  Repp,~J. Interface dipoles of Ir(ppy)3 on Cu(111). \emph{Nanoscale}
  \textbf{2019}, \emph{11}, 12695--12703\relax
\mciteBstWouldAddEndPuncttrue
\mciteSetBstMidEndSepPunct{\mcitedefaultmidpunct}
{\mcitedefaultendpunct}{\mcitedefaultseppunct}\relax
\EndOfBibitem
\bibitem[Ishii \latin{et~al.}(1999)Ishii, Sugiyama, Ito, and Seki]{Ishii1999}
Ishii,~H.; Sugiyama,~K.; Ito,~E.; Seki,~K. Energy Level Alignment and
  Interfacial Electronic Structures at Organic Metal and Organic Organic
  Interfaces. \emph{Adv. Mater.} \textbf{1999}, \emph{11}, 605--625\relax
\mciteBstWouldAddEndPuncttrue
\mciteSetBstMidEndSepPunct{\mcitedefaultmidpunct}
{\mcitedefaultendpunct}{\mcitedefaultseppunct}\relax
\EndOfBibitem
\bibitem[Braun \latin{et~al.}(2009)Braun, Salaneck, and Fahlman]{Braun2009}
Braun,~S.; Salaneck,~W.~R.; Fahlman,~M. Energy-Level Alignment at Organic/Metal
  and Organic/ Organic Interfaces. \emph{Adv. Mater.} \textbf{2009}, \emph{21},
  1450--1472\relax
\mciteBstWouldAddEndPuncttrue
\mciteSetBstMidEndSepPunct{\mcitedefaultmidpunct}
{\mcitedefaultendpunct}{\mcitedefaultseppunct}\relax
\EndOfBibitem
\bibitem[Torrente \latin{et~al.}(2008)Torrente, Franke, and
  Pascual]{Torrente2008}
Torrente,~I.~F.; Franke,~K.~J.; Pascual,~J.~I. Spectroscopy of C$_{60}$ single
  molecules: the role of screening on energy level alignment. \emph{J. Phys:
  Condens. Matt.} \textbf{2008}, \emph{20}, 184001\relax
\mciteBstWouldAddEndPuncttrue
\mciteSetBstMidEndSepPunct{\mcitedefaultmidpunct}
{\mcitedefaultendpunct}{\mcitedefaultseppunct}\relax
\EndOfBibitem
\bibitem[Repp \latin{et~al.}(2005)Repp, Meyer, Stojkovi\'{c}, Gourdon, and
  Joachim]{repp1}
Repp,~J.; Meyer,~G.; Stojkovi\'{c},~S.~M.; Gourdon,~A.; Joachim,~C. Molecules
  on {Insulating} {Films}: {Scanning}-{Tunneling} {Microscopy} {Imaging} of
  {Individual} {Molecular} {Orbitals}. \emph{Phys. Rev. Lett.} \textbf{2005},
  \emph{94}, 026803\relax
\mciteBstWouldAddEndPuncttrue
\mciteSetBstMidEndSepPunct{\mcitedefaultmidpunct}
{\mcitedefaultendpunct}{\mcitedefaultseppunct}\relax
\EndOfBibitem
\bibitem[Qiu \latin{et~al.}(2004)Qiu, Nazin, and Ho]{Qiu2004}
Qiu,~X.~H.; Nazin,~G.~V.; Ho,~W. Vibronic States in Single Molecule Electron
  Transport. \emph{Phys. Rev. Lett.} \textbf{2004}, \emph{92}, 206102\relax
\mciteBstWouldAddEndPuncttrue
\mciteSetBstMidEndSepPunct{\mcitedefaultmidpunct}
{\mcitedefaultendpunct}{\mcitedefaultseppunct}\relax
\EndOfBibitem
\bibitem[Kahle \latin{et~al.}(2012)Kahle, Deng, Malinowski, Tonnoir,
  Forment-Aliaga, Thontasen, Rinke, Le, Turkowski, Rahman, Rauschenbach,
  Ternes, and Kern]{Kahle2012}
Kahle,~S.; Deng,~Z.; Malinowski,~N.; Tonnoir,~C.; Forment-Aliaga,~A.;
  Thontasen,~N.; Rinke,~G.; Le,~D.; Turkowski,~V.; Rahman,~T.~S. \latin{et~al.}
   The Quantum Magnetism of Individual Manganese-12-Acetate Molecular Magnets
  Anchored at Surfaces. \emph{Nano Lett.} \textbf{2012}, \emph{12},
  518--521\relax
\mciteBstWouldAddEndPuncttrue
\mciteSetBstMidEndSepPunct{\mcitedefaultmidpunct}
{\mcitedefaultendpunct}{\mcitedefaultseppunct}\relax
\EndOfBibitem
\bibitem[Schulz \latin{et~al.}(2013)Schulz, Drost, H\"am\"al\"ainen, and
  Liljeroth]{Schulz2013}
Schulz,~F.; Drost,~R.; H\"am\"al\"ainen,~S.~K.; Liljeroth,~P. Templated
  self-assembly and local doping of molecules on epitaxial hexagonal boron
  nitride. \emph{ACS Nano} \textbf{2013}, \emph{7}, 11121--11128\relax
\mciteBstWouldAddEndPuncttrue
\mciteSetBstMidEndSepPunct{\mcitedefaultmidpunct}
{\mcitedefaultendpunct}{\mcitedefaultseppunct}\relax
\EndOfBibitem
\bibitem[Krane \latin{et~al.}(2018)Krane, Lotze, Reecht, Zhang, Briseno, and
  Franke]{Krane2018}
Krane,~N.; Lotze,~C.; Reecht,~G.; Zhang,~L.; Briseno,~A.~L.; Franke,~K.~J.
  High-Resolution Vibronic Spectra of Molecules on Molybdenum Disulfide Allow
  for Rotamer Identification. \emph{ACS Nano} \textbf{2018}, \emph{12},
  11698--11703\relax
\mciteBstWouldAddEndPuncttrue
\mciteSetBstMidEndSepPunct{\mcitedefaultmidpunct}
{\mcitedefaultendpunct}{\mcitedefaultseppunct}\relax
\EndOfBibitem
\bibitem[Mohn \latin{et~al.}(2012)Mohn, Gross, Moll, and Meyer]{mohn1}
Mohn,~F.; Gross,~L.; Moll,~N.; Meyer,~G. Imaging the charge distribution within
  a single molecule. \emph{Nature Nanotech.} \textbf{2012}, \emph{7},
  227--231\relax
\mciteBstWouldAddEndPuncttrue
\mciteSetBstMidEndSepPunct{\mcitedefaultmidpunct}
{\mcitedefaultendpunct}{\mcitedefaultseppunct}\relax
\EndOfBibitem
\bibitem[Kahn \latin{et~al.}(2003)Kahn, Koch, and Gao]{kahn1}
Kahn,~A.; Koch,~N.; Gao,~W. Electronic structure and electrical properties of
  interfaces between metals and pi-conjugated molecular films. \emph{J. Polym.
  Sci. B: Polym. Phys.} \textbf{2003}, \emph{41}, 2529--2548\relax
\mciteBstWouldAddEndPuncttrue
\mciteSetBstMidEndSepPunct{\mcitedefaultmidpunct}
{\mcitedefaultendpunct}{\mcitedefaultseppunct}\relax
\EndOfBibitem
\bibitem[Koch \latin{et~al.}(2005)Koch, Duhm, Rabe, Vollmer, and
  Johnson]{Koch2005}
Koch,~N.; Duhm,~S.; Rabe,~J.~P.; Vollmer,~A.; Johnson,~R.~L. Optimized Hole
  Injection with Strong Electron Acceptors at Organic-Metal Interfaces.
  \emph{Phys. Rev. Lett.} \textbf{2005}, \emph{95}, 237601\relax
\mciteBstWouldAddEndPuncttrue
\mciteSetBstMidEndSepPunct{\mcitedefaultmidpunct}
{\mcitedefaultendpunct}{\mcitedefaultseppunct}\relax
\EndOfBibitem
\bibitem[Nonnenmacher \latin{et~al.}(1991)Nonnenmacher, O'Boyle, and
  Wickramasinghe]{Nonnenmacher1991}
Nonnenmacher,~M.; O'Boyle,~M.~P.; Wickramasinghe,~H.~K. Kelvin probe force
  microscopy. \emph{Appl. Phys. Lett.} \textbf{1991}, \emph{58},
  2921--2923\relax
\mciteBstWouldAddEndPuncttrue
\mciteSetBstMidEndSepPunct{\mcitedefaultmidpunct}
{\mcitedefaultendpunct}{\mcitedefaultseppunct}\relax
\EndOfBibitem
\bibitem[Melitz \latin{et~al.}(2011)Melitz, Shen, Kummel, and Lee]{melitz1}
Melitz,~W.; Shen,~J.; Kummel,~A.~C.; Lee,~S. Kelvin probe force microscopy and
  its application. \emph{Surf. Sci. Rep.} \textbf{2011}, \emph{66}, 1--27\relax
\mciteBstWouldAddEndPuncttrue
\mciteSetBstMidEndSepPunct{\mcitedefaultmidpunct}
{\mcitedefaultendpunct}{\mcitedefaultseppunct}\relax
\EndOfBibitem
\bibitem[Gross \latin{et~al.}(2009)Gross, Mohn, Liljeroth, Repp, Giessibl, and
  Meyer]{Gross2009}
Gross,~L.; Mohn,~F.; Liljeroth,~P.; Repp,~J.; Giessibl,~F.~J.; Meyer,~G.
  Measuring the Charge State of an Adatom with Noncontact Atomic Force
  Microscopy. \emph{Science} \textbf{2009}, \emph{324}, 1428--1431\relax
\mciteBstWouldAddEndPuncttrue
\mciteSetBstMidEndSepPunct{\mcitedefaultmidpunct}
{\mcitedefaultendpunct}{\mcitedefaultseppunct}\relax
\EndOfBibitem
\bibitem[Schuler \latin{et~al.}(2014)Schuler, Liu, Geng, Decurtins, Meyer, and
  Gross]{schuler1}
Schuler,~B.; Liu,~S.-X.; Geng,~Y.; Decurtins,~S.; Meyer,~G.; Gross,~L. Contrast
  {Formation} in {Kelvin} {Probe} {Force} {Microscopy} of {Single} {{$\pi$}}
  {Conjugated} {Molecules}. \emph{Nano Lett.} \textbf{2014}, \emph{14},
  3342--3346\relax
\mciteBstWouldAddEndPuncttrue
\mciteSetBstMidEndSepPunct{\mcitedefaultmidpunct}
{\mcitedefaultendpunct}{\mcitedefaultseppunct}\relax
\EndOfBibitem
\bibitem[Albrecht \latin{et~al.}(2015)Albrecht, Repp, Fleischmann, Scheer,
  Ondr\'{a}\u{c}ek, and Jel\'{i}nek]{albrecht1}
Albrecht,~F.; Repp,~J.; Fleischmann,~M.; Scheer,~M.; Ondr\'{a}\u{c}ek,~M.;
  Jel\'{i}nek,~P. Probing {Charges} on the {Atomic} {Scale} by {Means} of
  {Atomic} {Force} {Microscopy}. \emph{Phys. Rev. Lett.} \textbf{2015},
  \emph{115}, 076101\relax
\mciteBstWouldAddEndPuncttrue
\mciteSetBstMidEndSepPunct{\mcitedefaultmidpunct}
{\mcitedefaultendpunct}{\mcitedefaultseppunct}\relax
\EndOfBibitem
\bibitem[Albrecht \latin{et~al.}(2015)Albrecht, Fleischmann, Scheer, Gross, and
  Repp]{Albrecht2015}
Albrecht,~F.; Fleischmann,~M.; Scheer,~M.; Gross,~L.; Repp,~J. Local tunneling
  decay length and Kelvin probe force spectroscopy. \emph{Phys. Rev. B}
  \textbf{2015}, \emph{92}, 235443\relax
\mciteBstWouldAddEndPuncttrue
\mciteSetBstMidEndSepPunct{\mcitedefaultmidpunct}
{\mcitedefaultendpunct}{\mcitedefaultseppunct}\relax
\EndOfBibitem
\bibitem[Krull \latin{et~al.}(2018)Krull, Castelli, Hapala, Kumar, Tadich,
  Capsoni, Edmonds, Hellerstedt, Burke, Jelinek, and Schiffrin]{krull2018}
Krull,~C.; Castelli,~M.; Hapala,~P.; Kumar,~D.; Tadich,~A.; Capsoni,~M.;
  Edmonds,~M.~T.; Hellerstedt,~J.; Burke,~S.~A.; Jelinek,~P. \latin{et~al.}
  Iron-based trinuclear metal-organic nanostructures on a surface with local
  charge accumulation. \emph{Nature Commun.} \textbf{2018}, \emph{9},
  3211\relax
\mciteBstWouldAddEndPuncttrue
\mciteSetBstMidEndSepPunct{\mcitedefaultmidpunct}
{\mcitedefaultendpunct}{\mcitedefaultseppunct}\relax
\EndOfBibitem
\bibitem[Rietsch \latin{et~al.}(2019)Rietsch, Witte, Sobottka, Germer, Becker,
  G\"uttler, Sarkar, Paulus, Resch-Genger, and Eigler]{rietsch1}
Rietsch,~P.; Witte,~F.; Sobottka,~S.; Germer,~G.; Becker,~A.; G\"uttler,~A.;
  Sarkar,~B.; Paulus,~B.; Resch-Genger,~U.; Eigler,~S. Diaminodicyanoquinones:
  Fluorescent Dyes with High Dipole Moments and Electron-Acceptor Properties.
  \emph{Angew. Chem. Int. Ed.} \textbf{2019}, \emph{58}, 8235--8239\relax
\mciteBstWouldAddEndPuncttrue
\mciteSetBstMidEndSepPunct{\mcitedefaultmidpunct}
{\mcitedefaultendpunct}{\mcitedefaultseppunct}\relax
\EndOfBibitem
\bibitem[Giessibl(2000)]{giessibl1}
Giessibl,~F.~J. Atomic resolution on {Si}-(111)(7x7) by noncontact atomic force
  microscopy with a force sensor based on a quartz tuning fork. \emph{Appl.
  Phys. Lett.} \textbf{2000}, \emph{76}, 1470--1472\relax
\mciteBstWouldAddEndPuncttrue
\mciteSetBstMidEndSepPunct{\mcitedefaultmidpunct}
{\mcitedefaultendpunct}{\mcitedefaultseppunct}\relax
\EndOfBibitem
\bibitem[Giannozzi \latin{et~al.}(2009)Giannozzi, Baroni, Bonini, Calandra,
  Car, Cavazzoni, Ceresoli, Chiarotti, Cococcioni, Dabo, Corso, de~Gironcoli,
  Fabris, Fratesi, Gebauer, Gerstmann, Gougoussis, Kokalj, Lazzeri,
  Martin-Samos, Marzari, Mauri, Mazzarello, Paolini, Pasquarello, Paulatto,
  Sbraccia, Scandolo, Sclauzero, Seitsonen, Smogunov, Umari, and
  Wentzcovitch]{Giannozzi2009}
Giannozzi,~P.; Baroni,~S.; Bonini,~N.; Calandra,~M.; Car,~R.; Cavazzoni,~C.;
  Ceresoli,~D.; Chiarotti,~G.~L.; Cococcioni,~M.; Dabo,~I. \latin{et~al.}
  {QUANTUM} {ESPRESSO}: a modular and open-source software project for quantum
  simulations of materials. \emph{J. Phys.: Condens. Matter} \textbf{2009},
  \emph{21}, 395502\relax
\mciteBstWouldAddEndPuncttrue
\mciteSetBstMidEndSepPunct{\mcitedefaultmidpunct}
{\mcitedefaultendpunct}{\mcitedefaultseppunct}\relax
\EndOfBibitem
\bibitem[Perdew \latin{et~al.}(1996)Perdew, Burke, and Ernzerhof]{Perdew1996}
Perdew,~J.~P.; Burke,~K.; Ernzerhof,~M. Generalized Gradient Approximation Made
  Simple. \emph{Phys. Rev. Lett.} \textbf{1996}, \emph{77}, 3865--3868\relax
\mciteBstWouldAddEndPuncttrue
\mciteSetBstMidEndSepPunct{\mcitedefaultmidpunct}
{\mcitedefaultendpunct}{\mcitedefaultseppunct}\relax
\EndOfBibitem
\bibitem[Vanderbilt(1990)]{Vanderbilt1990}
Vanderbilt,~D. Soft self-consistent pseudopotentials in a generalized
  eigenvalue formalism. \emph{Phys. Rev. B} \textbf{1990}, \emph{41},
  7892--7895\relax
\mciteBstWouldAddEndPuncttrue
\mciteSetBstMidEndSepPunct{\mcitedefaultmidpunct}
{\mcitedefaultendpunct}{\mcitedefaultseppunct}\relax
\EndOfBibitem
\bibitem[Grimme \latin{et~al.}(2010)Grimme, Antony, Ehrlich, and
  Krieg]{Grimme2010}
Grimme,~S.; Antony,~J.; Ehrlich,~S.; Krieg,~H. A consistent and accurate ab
  initio parametrization of density functional dispersion correction (DFT-D)
  for the 94 elements H-Pu. \emph{J. Chem. Phys.} \textbf{2010}, \emph{132},
  154104\relax
\mciteBstWouldAddEndPuncttrue
\mciteSetBstMidEndSepPunct{\mcitedefaultmidpunct}
{\mcitedefaultendpunct}{\mcitedefaultseppunct}\relax
\EndOfBibitem
\bibitem[Kachel \latin{et~al.}(2020)Kachel, Klein, Morbec, Sch\"oniger, Hutter,
  Schmid, Kratzer, Meyer, Tonner, and Gottfried]{Kachel2020}
Kachel,~S.~R.; Klein,~B.~P.; Morbec,~J.~M.; Sch\"oniger,~M.; Hutter,~M.;
  Schmid,~M.; Kratzer,~P.; Meyer,~B.; Tonner,~R.; Gottfried,~J.~M.
  Chemisorption and Physisorption at the Metal/Organic Interface: Bond Energies
  of Naphthalene and Azulene on Coinage Metal Surfaces. \emph{J. Phys. Chem. C}
  \textbf{2020}, \emph{124}, 8257--8268\relax
\mciteBstWouldAddEndPuncttrue
\mciteSetBstMidEndSepPunct{\mcitedefaultmidpunct}
{\mcitedefaultendpunct}{\mcitedefaultseppunct}\relax
\EndOfBibitem
\bibitem[Wolfram \latin{et~al.}(2022)Wolfram, Tariq, Fernandez, Muth, Gurrath,
  Wechsler, Franke, Williams, Steinr\"uck, Meyer, and Lytken]{Wolfram2022}
Wolfram,~A.; Tariq,~Q.; Fernandez,~C.~C.; Muth,~M.; Gurrath,~M.; Wechsler,~D.;
  Franke,~M.; Williams,~F.~J.; Steinr\"uck,~H.-P.; Meyer,~B. \latin{et~al.}
  Adsorption energies of porphyrins on MgO(100): An experimental benchmark for
  dispersion-corrected density-functional theory. \emph{Surf. Sci.}
  \textbf{2022}, \emph{717}, 121979\relax
\mciteBstWouldAddEndPuncttrue
\mciteSetBstMidEndSepPunct{\mcitedefaultmidpunct}
{\mcitedefaultendpunct}{\mcitedefaultseppunct}\relax
\EndOfBibitem
\bibitem[Tersoff(1985)]{TH85}
Tersoff,~J.; Hamann,~D.~R. Theory of the scanning tunneling microscope.
  \emph{Phys. Rev. B} \textbf{1985}, \emph{31}, 805--813\relax
\mciteBstWouldAddEndPuncttrue
\mciteSetBstMidEndSepPunct{\mcitedefaultmidpunct}
{\mcitedefaultendpunct}{\mcitedefaultseppunct}\relax
\EndOfBibitem
\bibitem[van~der Lit \latin{et~al.}(2016)van~der Lit, Di~Cicco, Hapala,
  Jelinek, and Swart]{vanderlit1}
van~der Lit,~J.; Di~Cicco,~F.; Hapala,~P.; Jelinek,~P.; Swart,~I. Submolecular
  {Resolution} {Imaging} of {Molecules} by {Atomic} {Force} {Microscopy}: {The}
  {Influence} of the {Electrostatic} {Force}. \emph{Phys. Rev. Lett.}
  \textbf{2016}, \emph{116}, 096102\relax
\mciteBstWouldAddEndPuncttrue
\mciteSetBstMidEndSepPunct{\mcitedefaultmidpunct}
{\mcitedefaultendpunct}{\mcitedefaultseppunct}\relax
\EndOfBibitem
\bibitem[Pavli\u{c}ek \latin{et~al.}(2013)Pavli\u{c}ek, Swart, Niedenf\"uhr,
  Meyer, and Repp]{pavlicek1}
Pavli\u{c}ek,~N.; Swart,~I.; Niedenf\"uhr,~J.; Meyer,~G.; Repp,~J. Symmetry
  {Dependence} of {Vibration}-{Assisted} {Tunneling}. \emph{Phys. Rev. Lett.}
  \textbf{2013}, \emph{110}, 136101\relax
\mciteBstWouldAddEndPuncttrue
\mciteSetBstMidEndSepPunct{\mcitedefaultmidpunct}
{\mcitedefaultendpunct}{\mcitedefaultseppunct}\relax
\EndOfBibitem
\bibitem[Reecht \latin{et~al.}(2020)Reecht, Krane, Lotze, Zhang, Briseno, and
  Franke]{Reecht2020}
Reecht,~G.; Krane,~N.; Lotze,~C.; Zhang,~L.; Briseno,~A.~L.; Franke,~K.~J.
  Vibrational Excitation Mechanism in Tunneling Spectroscopy beyond the
  Franck-Condon Model. \emph{Phys. Rev. Lett.} \textbf{2020}, \emph{124},
  116804\relax
\mciteBstWouldAddEndPuncttrue
\mciteSetBstMidEndSepPunct{\mcitedefaultmidpunct}
{\mcitedefaultendpunct}{\mcitedefaultseppunct}\relax
\EndOfBibitem
\bibitem[Heimel \latin{et~al.}(2008)Heimel, Romaner, Zojer, and
  Bredas]{heimel1}
Heimel,~G.; Romaner,~L.; Zojer,~E.; Bredas,~J.-L. The {Interface} {Energetics}
  of {Self} {Assembled} {Monolayers} on {Metals}. \emph{Accounts of Chemical
  Research} \textbf{2008}, \emph{41}, 721--729\relax
\mciteBstWouldAddEndPuncttrue
\mciteSetBstMidEndSepPunct{\mcitedefaultmidpunct}
{\mcitedefaultendpunct}{\mcitedefaultseppunct}\relax
\EndOfBibitem
\bibitem[Koch \latin{et~al.}(2002)Koch, Kahn, Ghijsen, Pireaux, Schwartz,
  Johnson, and Elschner]{koch1}
Koch,~N.; Kahn,~A.; Ghijsen,~J.; Pireaux,~J.-J.; Schwartz,~J.; Johnson,~R.~L.;
  Elschner,~A. Conjugated organic molecules on metal versus polymer electrodes:
  {Demonstration} of a key energy level alignment mechanism. \emph{Appl. Phys.
  Lett.} \textbf{2002}, \emph{82}, 70--72\relax
\mciteBstWouldAddEndPuncttrue
\mciteSetBstMidEndSepPunct{\mcitedefaultmidpunct}
{\mcitedefaultendpunct}{\mcitedefaultseppunct}\relax
\EndOfBibitem
\bibitem[Schlesinger(2017)]{schlesinger1}
Schlesinger,~R. \emph{Energy-Level Control at Hybrid Inorganic/Organic
  Semiconductor Interfaces}; Springer International Publishing, 2017\relax
\mciteBstWouldAddEndPuncttrue
\mciteSetBstMidEndSepPunct{\mcitedefaultmidpunct}
{\mcitedefaultendpunct}{\mcitedefaultseppunct}\relax
\EndOfBibitem
\end{mcitethebibliography}

\providecommand{\latin}[1]{#1}
\makeatletter
\providecommand{\doi}
  {\begingroup\let\do\@makeother\dospecials
  \catcode`\{=1 \catcode`\}=2 \doi@aux}
\providecommand{\doi@aux}[1]{\endgroup\texttt{#1}}
\makeatother
\providecommand*\mcitethebibliography{\thebibliography}
\csname @ifundefined\endcsname{endmcitethebibliography}
  {\let\endmcitethebibliography\endthebibliography}{}

\end{document}